\documentclass[conference]{IEEEtran}

\ifCLASSOPTIONcompsoc

  \usepackage[nocompress]{cite}
  
\else

  \usepackage{cite}
\fi

\makeatletter
\def\endthebibliography{%
  \def\@noitemerr{\@latex@warning{Empty `thebibliography' environment}}%
  \endlist
}

\usepackage{graphicx}
\usepackage{subfigure}
\usepackage{setspace}
\usepackage{caption}
\usepackage{float}
\usepackage{booktabs}
\usepackage[font=small,skip=0.8pt]{caption}
\usepackage{hyperref}
\usepackage{arabtex}
\usepackage{utf8}
\setcode{utf8}
\usepackage{placeins}
\usepackage{ragged2e}
\usepackage{footnote}
\usepackage{enumitem}
\setitemize{noitemsep,topsep=0pt,parsep=0pt,partopsep=0pt}
\makesavenoteenv{table}
\makesavenoteenv{tabular}
\begin{document}
\title{Detecting Requirements Smells With Deep Learning: Experiences,  Challenges and Future Work}
\author{
  \IEEEauthorblockN{Mohammad Kasra Habib}
  \IEEEauthorblockA{\textit{Institute of Software Engineering} \\
    \textit{University of Stuttgart}\\
    Stuttgart,  Germany \\
    kasra.habib@iste.uni-stuttgart.de}
  \and
  \IEEEauthorblockN{Stefan Wagner}
   \IEEEauthorblockA{\textit{Institute of Software Engineering} \\
    \textit{University of Stuttgart}\\
    Stuttgart,  Germany \\
    stefan.wagner@iste.uni-stuttgart.de}
    \and
    \IEEEauthorblockN{Daniel Graziotin}
   \IEEEauthorblockA{\textit{Institute of Software Engineering} \\
    \textit{University of Stuttgart}\\
    Stuttgart,  Germany \\
    daniel.graziotin@iste.uni-stuttgart.de}
    }

\IEEEtitleabstractindextext{
\begin{abstract}
\justifying
Requirements Engineering (RE) is one of the initial phases when building a software system. The success or failure of a software project is firmly tied to this phase, based on communication among stakeholders using natural language. The problem with natural language is that it can easily lead to different understandings if it is not expressed precisely by the stakeholders involved. This results in building a product which is different from the expected one.  Previous work proposed to enhance the quality of the software requirements by detecting language errors based on ISO 29148 requirements language criteria. The existing solutions apply classical Natural Language Processing (NLP) to detect them. NLP has some limitations, such as domain dependability which results in poor generalization capability.  Therefore, this work aims to improve the previous work by creating a manually labeled dataset and using ensemble learning, Deep Learning (DL), and techniques such as word embeddings and transfer learning to overcome the generalization problem that is tied with classical NLP and improve precision and recall metrics using a manually labeled dataset. The current findings show that the dataset is unbalanced and which class examples should be added more. It is tempting to train algorithms even if the dataset is not considerably representative.  Whence, the results show that models are overfitting; in Machine Learning this issue is adressed by adding more instances to the dataset, improving label quality,  removing noise, and reducing the learning algorithms complexity, which is planned for this research.
\end{abstract}

}

\maketitle

\IEEEdisplaynontitleabstractindextext

\IEEEpeerreviewmaketitle

\vspace{1.3cm}
\IEEEraisesectionheading{\section{Introduction}\label{sec:introduction}}
\IEEEPARstart{A}{} crucial activity in the software development life cycle is RE \cite{pohl2010requirements,  nuseibeh2000requirements,  berry1995importance}.  RE involves three key activities, requirements elicitation, specification, and validation \cite{sommerville2015software, iqbal2018bird}.  This research focus on improving the requirements specification activity, and improvement in requirements itself mean making them fathomable; therefore, this can be helpful for the validation activity too. %
Requirements for the software to be developed are expressed in natural language by the stakeholders who have limited or lack knowledge in the software engineering domain \cite{dalpiaz_natural_2018, nguyen2012reindetector,  iqbal2018bird,  ferrari2017towards}; Sommerville  \cite{sommerville2015software} states that \lq\lq customers for a system often find it difficult to translate their goals into measurable requirements,\rq\rq\ which results in the requirements being taint with ambiguities and general terms \cite{selvyanti2017requirements}. Therefore,  \cite{sommerville2015software} proposes that the requirements should be rewritten quantitively to test them objectively.   For example, ISO 29148 suggests words to be avoided in software requirements \cite{selvyanti2017requirements}: \textit{superlatives},  \textit{subjective language}, \textit{vague pronouns}, \textit{ ambiguous adverbs and adjectives},  \textit{open-ended and non-verifiable terms}, \textit{comparatives phrases},  \textit{loopholes},  \textit{incomplete references},  and \textit{negative statements}.

Imprecise requirement specifications can lead to disputes between software engineers and customers because they can be misinterpreted \cite{sommerville2015software}. Software requirements that are ambiguous and are expressed in general terms heavily cost the software project both in terms of time and budget \cite{iqbal2018bird}. 

One approach for validating requirements to support quality management and reduce time and budget can be manual inspection to follow the language criteria prescribed within ISO  29148 to determine the defects in the requirements \cite{selvyanti2017requirements}. 

Another approach would be to build a tool to enhance requirements quality automatically.  Although there is limited research in this specific area \cite{pekar_improvement_2014},  \cite{femmer2017rapid} and \cite{rosadini2017using} are reasonable attempts using NLP in this direction. However, both studies are presenting a low precision and recall. Moreover, both report a poor generalization capability of the approach over the domain as the most significant limitation of their work.

Nevertheless, a more robust approach to detect requirements smells\footnote{The term \textit{requirements smells} is introduced by Femmer et al.  \cite{femmer2017rapid}; this refers to a quality violation in requirements. This paper picks this term to refer to defects in requirements.} and precise method to this challenge is combining Deep Learning with classical NLP since Deep Learning is recognized for greater generalization and solid prediction capability, which has recently proven successful for binary or multi-class requirements classification \cite{hey2020norbert, subedi2021application, weiss2016survey,  9402628, otter2020survey, li2017deep, young2018recent}.  Since tackling this challenge with machine learning means multi-class multi-label classification,  Deep Learning should perform outstanding in the presence of a rich dataset as well.

Therefore, this research plans to create a rich, manually labeled dataset; each language criterion defined in ISO 29148 is considered a label, and deploying Deep Learning to improve the previously observed precision and recall in \cite{femmer2017rapid,  rosadini2017using},  which increases practitioners' trust in using this tool.  Once a Deep Learning model is trained, it usually makes better predictions than classical Machine Learning approaches or classical NLP \cite{geron2019hands}, which could overcome the poor generalization capability of \cite{femmer2017rapid, rosadini2017using} since these models are less sensitive to domain. Besides, most recent studies reveal that applying transfer learning which is a technique used in Deep Learning, is successful since RE suffers from a lack of datasets \cite{hey2020norbert, subedi2021application, weiss2016survey,  9402628, otter2020survey, li2017deep, young2018recent, torrey2010transfer}.  Furthermore, this work plans to build an ensemble of already trained models, which has been helpful for other domains \cite{webb2004multistrategy, fern2003online, sewell2008ensemble} and will be done for the first time in this research for RE.

\section{Related Work}
Various research efforts exist to enhance quality for software requirements; some focus on practicing classical NLP, such as \cite{femmer2017rapid,  rosadini2017using},  where others use Artificial Intelligence (AI) \cite{ferrari2012using, yang2011analysing}.  We regard \cite{femmer2017rapid,  rosadini2017using} closer to our work, although they use NLP approaches, whereas \cite{ferrari2012using, yang2011analysing} employ AI approaches. Additionally,  \cite{femmer2017rapid,  rosadini2017using} are recently published and focus on almost the same criteria for requirements smells.

Rosadini et al.  \cite{rosadini2017using} investigate to what extent NLP can be practically applied to detect defects in requirements in the railway domain. They develop a dataset manually annotated by domain experts. They reach an average of 85.6\% precision on the domain requirements dataset. The paper concludes that it is essential to develop the tools tailored for the patterns specific to the company's needs and that the NLP tools should be used by requirements editors to limit the amount of poor writing style. One of the limitations of the study is the significant amount of false positives generated by the NLP-based approach, which is planned to be addressed by further adjustments in the future. Nevertheless, they say, \lq\lq NLP is part of the solution.\rq\rq

Another significant milestone in this direction is a paper \cite{femmer2017rapid} by Femmer et al.  that also applies NLP to detect requirements smells based on ISO 29148 language criteria.  This approach yields an average precision of 59\% and recall of 82\%. Moreover,  Femmer et al. \cite{femmer2017rapid} conclude that interviewed practitioners agreed on the usefulness of the developed tool.  Additionally,  the authors report that the practitioners have different views on integrating this tool in the quality assurance process.  However,  all the interviewed practitioners state that the requirements smell tool can be used by the person writing requirements as a support, not as the primary tool for requirements quality check.  The achieved recall is likely the culprit.  A lower recall means categorizing faulty requirements (true positive) as non-faulty requirements (false negative); for the person responsible for writing requirements, this means almost the same effort to search for those faulty requirements.

Both \cite{femmer2017rapid, rosadini2017using} are solid studies that highlight the importance of quality assurance for RE.  Although all the cited efforts are in a way or another related to the aims of this research, we see issues.  For example, \cite{femmer2017rapid,  rosadini2017using} suffer from limitations;  they have a poor generalization capability and present a low precision and recall.  We argue that this is heavily tied with the use of classical NLP.  Most NLP approaches are domain-dependent \cite{khan2016survey, vogelsang_automatic_2021, asghar2016automatic}; as tool decisions are often based on some set of rules,  rather than learning like Deep Learning approaches and if they do not match, it fails.  Therefore, our research takes a step forward in improving preceding works' performance metrics and breaking their limitations by leveraging Deep Learning, Machine Learning, and NLP.

\section{Planned Approach} 
We aim to apply Deep Learning to improve requirements smell detection and pursue the following research questions (RQ): 
\begin{itemize}
\item \textbf{RQ1}: Is Deep Learning improving NLP-based solutions for requirements smell detection? 
\item \textbf{RQ2}: To what degree does Deep Learning improve classical NLP's generalization disability for requirements smells' detection?
\item \textbf{RQ3}: Do pre-trained models offer a better performance on a small size requirements dataset with multi-labels and multi-classes?
\item \textbf{RQ4}: To what degree does data from closed-source projects influence the generalization capability?
\item \textbf{RQ5}: Can the Deep Learning-based requirements smell detection tool be trusted as a primary quality assurance tool during RE?
\end{itemize}
RQ 5 is qualitative and will be addressed once questions 1 -- 4 is fulfilled.  For RQ 1--4 we plan to build a customized preprocessing pipeline to clean and represent requirements as a set of word embedding features. Then we deploy Deep Learning, using transfer learning which is suitable when low-quality data is available.  The selected models for transfer learning are BERT \cite{devlin2018bert}, fastText,\cite{joulin2016fasttext} and ELMo \cite{buyukoz2020analyzing}; since recently they have shown promising performance for alike tasks \cite{vogelsang_automatic_2021,  hey2020norbert, subedi2021application, weiss2016survey,  9402628, otter2020survey, li2017deep, young2018recent, torrey2010transfer}.  Eventually, we aim to accomplish ensemble learning. Based on our experience with machine learning, combining weak learners with strong learners makes ensemble learning solid. SVM and Naive Bayes will be trained to put together with the Deep Learning models for ensemble learning.
The current challenge is a multi-class multi-label classification task since one requirement can be taint with different defects simultaneously; this implies instances to be classified in several classes. 

Multi-class multi-label classification is a complex task requiring further research \cite{shi2014drift} since most studies are dedicated to binary classification.  Pant et al.  \cite{pant2019multi} report that \textit{handling dimensionality},  \textit{data cleaning},  \textit{label dependency},  \textit{label uncertainty},  \textit{label drifting},  and \textit{data imbalance} are the challenges to be tackled for multi-class multi-label classification. This paper finds very little support for multi-class multi-label classification tasks provided by machine learning API based on initial experiments and results.  To solve this issue, researchers either have to develop tools from scratch or tweak the current machine learning APIs to provide support. Another hurdle for multi-class multi-label classification is the lack of a rich dataset specifically in the RE domain; training Deep Learning models requires showing many instances to reach solid generalization capability. Therefore, this study proposes adding support in machine learning APIs and rich datasets to the list of challenges for multi-class multi-label classification.

A gap in employing deep learning models to the RE domain is lack of datasets \cite{hey2020norbert,  chawla2002smote,  dalpiaz_natural_2018}. Hence, to make this work thriving, an important milestone is to assemble requirements for creating a dataset that satisfies the task. We gathered requirements specification documents from 28 closed-source projects.  We extract 454 requirements from closed-source projects and 2792 requirements from the PURE dataset \cite{ferrari2017pure}.  Our work intends to expand the dataset size to around 15000--20000 instances.

\section{Current Results}
Our effort combines requirements from all projects to get an initial insight about the currently collected requirements,  which results in a dataset size of 3246. Then, we adopt a technique called self-supervised learning to generate labels for each instance automatically. Self-supervised learning is the application of generating labels from  data \cite{geron2019hands}. The ultimate goal of this study is to carry manual labeling once data collection is complete, not to use self-supervised learning. Nevertheless, to achieve a quick insight,  NLP is applied as previous studies for each sample from the culled dataset to capture five smells: subjective language (JJ), ambiguous adverbs and adjectives (RB), superlatives (JJS), comparatives (JJR), and vague pronouns (WDT).

\begin{figure}[htb]
\centering
\begin{minipage}{.5\textwidth}
\raggedright
\subfigure[Class distribution]{\includegraphics[width = 4.5cm]{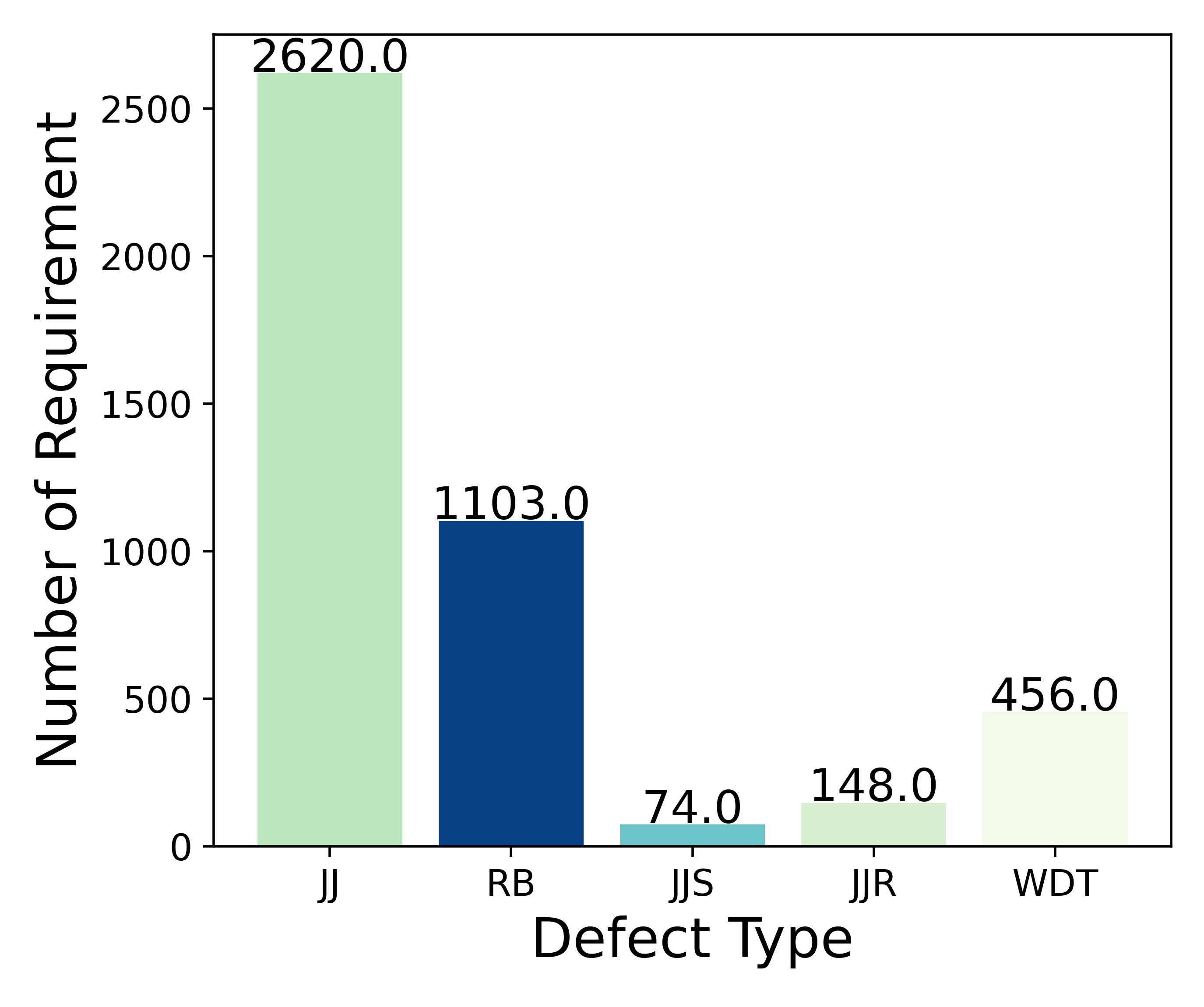}}
\end{minipage}
\begin{minipage}{.5\textwidth}
\raggedleft
\vspace{-4.4cm}
\subfigure[Requirements with multiple label]{\includegraphics[width = 4.5cm]{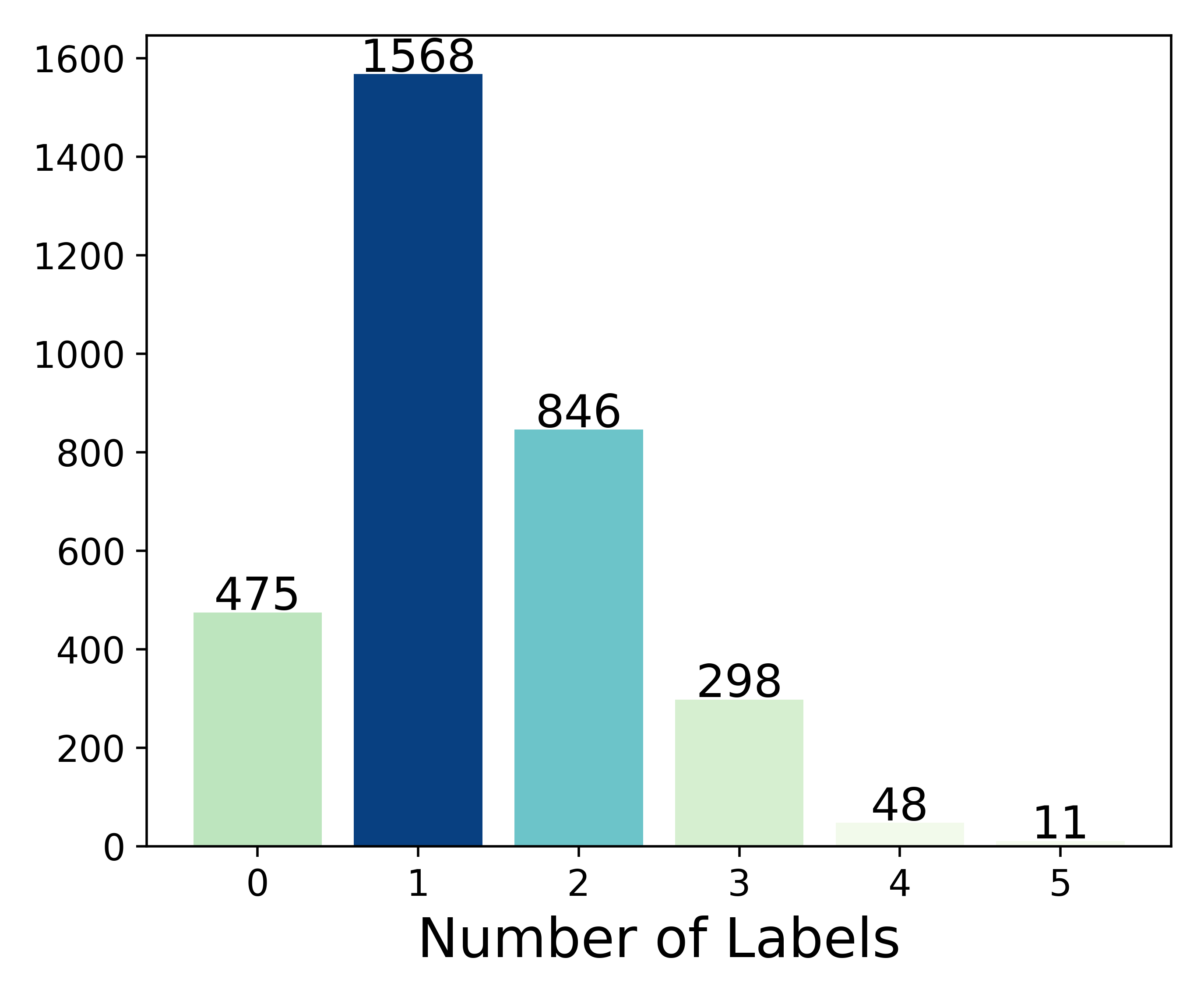}}
\end{minipage}
\caption{Current dataset's class and label distribution}
\end{figure}

Self-supervised learning does not fit perfectly in this context. However, some game-changing information can be retrieved at this stage; the insights shifts this work towards the best direction. From Fig. 1, it is easy to construe two critical features to be considered while developing the final dataset; subplot (a) and (b). 

Subplot (a) depicts frequencies per requirement category. It can be seen that the dataset is highly unbalanced. Therefore, more instances with superlatives, comparatives, and vague pronouns must be added to the final dataset to correct the classifier's shun bias predictions for dominant classes.  Furthermore,  one can observe that most of the requirements are taint with \textit{subjective language (JJ)}, and \textit{ambiguous adverbs and adjective (RB)} smells. Besides,  subplot (b) shows the requirements having multiple labels. As it is evident, most of the instances either have one or two labels. It can be inferred that the dataset is less inclined to label drifting; this is a good thing to have \lq\lq since the interest in labels starts drifting as it is hidden conceptually\rq\rq\ \cite{pant2019multi}, which makes the decision making harder. Yet,   \textit{let us} not celebrate, i.e., this is a small portion of the dataset with poorly labeled instances.

Even though the data quality is not considered at this stage, it is tempting to run a few basic machine learning algorithms to observe their performance. Therefore, the dataset is fed to a direct preprocessing pipeline, and then they are converted to their vector representation applying \textit{word-1-grams} and \textit{TFIDF}. Next, Synthetic Minority Oversampling Technique (SMOTE)\footnote{This type of data augmentation usually used with imbalance dataset to prevent accuracy paradox and overfitting \cite{chawla2002smote}.} \cite{chawla2002smote} is employed to balance the class distributions dataset. The chosen models are Multilayer Perceptron (MLP) and SVM. Next, algorithm hyperparameters are fine-tuned and trained with five-fold cross-validation.

\begin{table}[h]
\centering
\caption{Models performance measurement}
\begin{tabular}{@{}lccc@{}}
\toprule
             & \multicolumn{1}{l}{\small\textbf{Precison}} & \multicolumn{1}{l}{\small\textbf{Recall}} & \multicolumn{1}{l}{\small\textbf{$f_1$}} \\ \midrule
\small\textbf{MLP} & 0.87                                  & 0.79                                & 0.83                               \\
\small\textbf{SVM} & 0.89                                  & 0.81                                & 0.84                               \\ \bottomrule
\end{tabular}
\end{table}

From Table I, both models show a considerable performance. However, since this is a multi-class multi-label classification problem which is different from binary classification, directly looking at the precision, recall and $f_1$ score, or calculating loss can be deceiving. One approach is to check model performance with hamming loss, but this is just a number and not intuitive. Therefore, average learning curves over all classes for each algorithm are plotted (Fig. 2) to show how the algorithms perform during training and validation.

\begin{figure}[htb]
\includegraphics[width = 9cm]{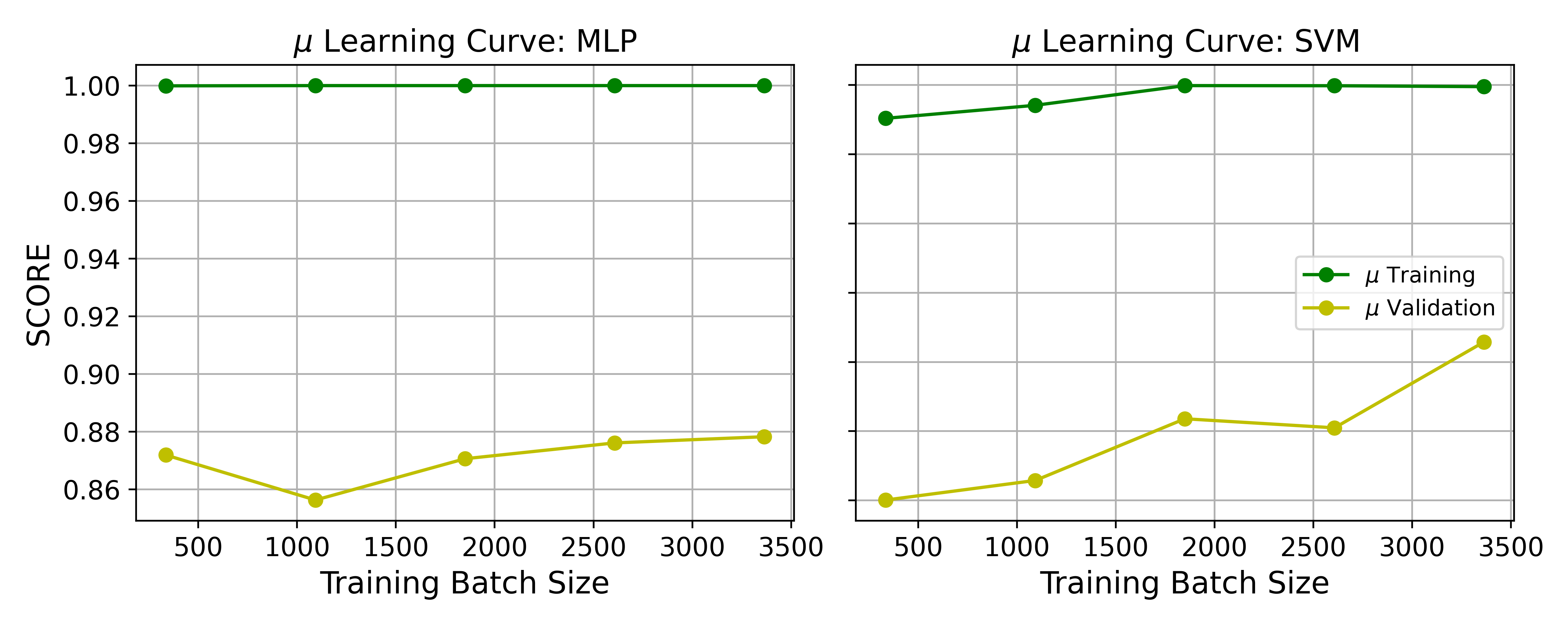}
\caption{Model's learning curves}
\end{figure}

From Fig. 2, one can easily pick that both models are overfitting (surprise!). There is a big gap between each' two curves. Both models perform significantly better on the training set (green curve) than the validation set (yellow curve). When a model is overfitting, it is either due to poor data quality or if the model is complex. Our selected models are not complex with a lot of hyperparameters to reduce them, implying that the problem is with the dataset since it suffers from insufficient instances, aberrant labeling, and noise. 

\section{Conclusion and Future Work}
This paper presents the importance of requirements for smell detection. Then, it proposes solutions based on Deep Learning to address the limitations developed by applying pure Natural Language Processing. Our work is at its early stage; it describes the building process and experience learned up until this point. The achievements in this effort are so far promising. Some of the findings can be used for other studies. To the best of our knowledge, this is the first study of multi-class multi-label classification using Deep Learning for RE, which already suggests API support for multi-class multi-label to build and lack of requirements dataset with multi-label instances.

As this is ongoing research, our current efforts are to build a dataset to address the lack of a dataset for this purpose which also solves the current overfitting problem. Once all requirements are collected from donated software requirements specifications, experts will label them manually. Then, a customized input pipeline will be developed to clean and represent requirements as a set of word embedding features. Ultimately, an ensemble of already trained models will be deployed to fit it for the data. Recently, transfer learning has proven to be the best tool in the RE domain since the lack of dataset, or miniature size datasets is only available. The expected benefits include a robust model with higher precision and recall metrics with a better generalization capability over the domain.

\bibliographystyle{IEEEtran}
\bibliography{./bib/references}

\begin{thebibliography}{10}
\providecommand{\url}[1]{#1}
\csname url@samestyle\endcsname
\providecommand{\newblock}{\relax}
\providecommand{\bibinfo}[2]{#2}
\providecommand{\BIBentrySTDinterwordspacing}{\spaceskip=0pt\relax}
\providecommand{\BIBentryALTinterwordstretchfactor}{4}
\providecommand{\BIBentryALTinterwordspacing}{\spaceskip=\fontdimen2\font plus
\BIBentryALTinterwordstretchfactor\fontdimen3\font minus
  \fontdimen4\font\relax}
\providecommand{\BIBforeignlanguage}[2]{{%
\expandafter\ifx\csname l@#1\endcsname\relax
\typeout{** WARNING: IEEEtran.bst: No hyphenation pattern has been}%
\typeout{** loaded for the language `#1'. Using the pattern for}%
\typeout{** the default language instead.}%
\else
\language=\csname l@#1\endcsname
\fi
#2}}
\providecommand{\BIBdecl}{\relax}
\BIBdecl

\bibitem{pohl2010requirements}
L.~A. Macaulay, \emph{Requirements engineering}.\hskip 1em plus 0.5em minus
  0.4em\relax Springer Science \& Business Media, 2012.

\bibitem{nuseibeh2000requirements}
B.~Nuseibeh and S.~Easterbrook, ``Requirements engineering: a roadmap,'' in
  \emph{Proceedings of the {Conference} on {The} {Future} of {Software}
  {Engineering}}.\hskip 1em plus 0.5em minus 0.4em\relax ACM, 2000, pp. 35--46.

\bibitem{berry1995importance}
D.~M. Berry, ``\BIBforeignlanguage{en}{The importance of ignorance in
  requirements engineering},'' \emph{\BIBforeignlanguage{en}{Journal of Systems
  and Software}}, vol.~28, no.~2, pp. 179--184, Feb. 1995.

\bibitem{sommerville2015software}
I.~Sommerville, ``\BIBforeignlanguage{en}{Software engineering}.''\hskip 1em
  plus 0.5em minus 0.4em\relax Pearson Education, 2016.

\bibitem{iqbal2018bird}
T.~Iqbal, P.~Elahidoost, and L.~Lucio, ``A {Bird}'s {Eye} {View} on
  {Requirements} {Engineering} and {Machine} {Learning},'' in \emph{2018 25th
  {Asia}-{Pacific} {Software} {Engineering} {Conference} ({APSEC})}.\hskip 1em
  plus 0.5em minus 0.4em\relax IEEE, 2018, pp. 11--20.

\bibitem{dalpiaz_natural_2018}
F.~Dalpiaz, A.~Ferrari, X.~Franch, and C.~Palomares, ``Natural {Language}
  {Processing} for {Requirements} {Engineering}: {The} {Best} {Is} {Yet} to
  {Come},'' \emph{IEEE Software}, vol.~35, no.~5, pp. 115--119, Sep. 2018.

\bibitem{nguyen2012reindetector}
T.~H. Nguyen, B.~Q. Vo, M.~Lumpe, and J.~Grundy, ``{REInDetector}: a framework
  for knowledge-based requirements engineering,'' in \emph{Proceedings of the
  27th {IEEE}/{ACM} international conference on automated software
  engineering}.\hskip 1em plus 0.5em minus 0.4em\relax IEEE/ACM, 2012, pp.
  386--389.

\bibitem{ferrari2017towards}
A.~Ferrari, G.~O. Spagnolo, and S.~Gnesi, ``Towards a {Dataset} for {Natural}
  {Language} {Requirements} {Processing}.'' in \emph{{REFSQ}
  {Workshops}}.\hskip 1em plus 0.5em minus 0.4em\relax CEUR, 2017.

\bibitem{selvyanti2017requirements}
\BIBentryALTinterwordspacing
ISO/IEC/IEEE, ``\BIBforeignlanguage{en}{Systems and software engineering --
  life cycle processes -- requirements engineering}.'' [Online]. Available:
  \url{https://www.iso.org/cms/render/live/en/sites/isoorg/contents/data/standard/07/20/72089.html}
\BIBentrySTDinterwordspacing

\bibitem{pekar_improvement_2014}
V.~Pekar, M.~Felderer, and R.~Breu, ``Improvement {Methods} for {Software}
  {Requirement} {Specifications}: {A} {Mapping} {Study},'' in \emph{2014 9th
  {International} {Conference} on the {Quality} of {Information} and
  {Communications} {Technology}}, Sep. 2014, pp. 242--245.

\bibitem{femmer2017rapid}
H.~Femmer, D.~M. Fernández, S.~Wagner, and S.~Eder, ``Rapid quality assurance
  with requirements smells,'' \emph{Journal of Systems and Software}, vol. 123,
  pp. 190--213, 2017.

\bibitem{rosadini2017using}
B.~Rosadini, A.~Ferrari, G.~Gori, A.~Fantechi, S.~Gnesi, I.~Trotta, and
  S.~Bacherini, ``\BIBforeignlanguage{en}{Using {NLP} to {Detect}
  {Requirements} {Defects}: {An} {Industrial} {Experience} in the {Railway}
  {Domain}},'' in \emph{\BIBforeignlanguage{en}{Requirements {Engineering}:
  {Foundation} for {Software} {Quality}}}, ser. Lecture {Notes} in {Computer}
  {Science}, P.~Grünbacher and A.~Perini, Eds.\hskip 1em plus 0.5em minus
  0.4em\relax Springer International Publishing, 2017, pp. 344--360.

\bibitem{hey2020norbert}
T.~Hey, J.~Keim, A.~Koziolek, and W.~F. Tichy, ``{NoRBERT}: {Transfer} learning
  for requirements classification,'' in \emph{2020 {IEEE} 28th {International}
  {Requirements} {Engineering} {Conference} ({RE})}.\hskip 1em plus 0.5em minus
  0.4em\relax IEEE, 2020, pp. 169--179.

\bibitem{subedi2021application}
I.~M. Subedi, M.~Singh, V.~Ramasamy, and G.~S. Walia, ``Application of
  back-translation: a transfer learning approach to identify ambiguous software
  requirements,'' in \emph{Proceedings of the 2021 {ACM} {Southeast}
  {Conference}}.\hskip 1em plus 0.5em minus 0.4em\relax ACM, 2021, pp.
  130--137.

\bibitem{weiss2016survey}
K.~Weiss, T.~M. Khoshgoftaar, and D.~Wang, ``A survey of transfer learning,''
  \emph{Journal of Big data}, vol.~3, no.~1, pp. 1--40, 2016.

\bibitem{9402628}
A.~Ferrari, L.~Zhao, and W.~Alhoshan, ``{NLP} for {Requirements} {Engineering}:
  {Tasks}, {Techniques}, {Tools}, and {Technologies},'' in \emph{2021
  {IEEE}/{ACM} 43rd {International} {Conference} on {Software} {Engineering}:
  {Companion} {Proceedings} ({ICSE}-{Companion})}.\hskip 1em plus 0.5em minus
  0.4em\relax IEEE, 2021, pp. 322--323.

\bibitem{otter2020survey}
D.~W. Otter, J.~R. Medina, and J.~K. Kalita, ``A survey of the usages of deep
  learning for natural language processing,'' \emph{IEEE Transactions on Neural
  Networks and Learning Systems}, 2020.

\bibitem{li2017deep}
H.~Li, ``Deep learning for natural language processing: advantages and
  challenges,'' \emph{National Science Review}, vol.~5, no.~1, pp. 24--26,
  2017.

\bibitem{young2018recent}
T.~Young, D.~Hazarika, S.~Poria, and E.~Cambria, ``Recent trends in deep
  learning based natural language processing,'' \emph{IEEE Computational
  intelligence magazine}, vol.~13, no.~3, pp. 55--75, 2018.

\bibitem{geron2019hands}
A.~Géron, \emph{Hands-on machine learning with {Scikit}-{Learn}, {Keras}, and
  {TensorFlow}: {Concepts}, tools, and techniques to build intelligent
  systems}.\hskip 1em plus 0.5em minus 0.4em\relax O'Reilly Media, 2019.

\bibitem{torrey2010transfer}
L.~Torrey and J.~Shavlik, ``Transfer learning,'' in \emph{Handbook of research
  on machine learning applications and trends: algorithms, methods, and
  techniques}.\hskip 1em plus 0.5em minus 0.4em\relax IGI global, 2010, pp.
  242--264.

\bibitem{webb2004multistrategy}
G.~Webb and Z.~Zheng, ``Multistrategy ensemble learning: reducing error by
  combining ensemble learning techniques,'' \emph{IEEE Transactions on
  Knowledge and Data Engineering}, vol.~16, no.~8, pp. 980--991, Aug. 2004.

\bibitem{fern2003online}
\BIBentryALTinterwordspacing
A.~Fern and R.~Givan, ``\BIBforeignlanguage{en}{Online {Ensemble} {Learning}:
  {An} {Empirical} {Study}},'' \emph{\BIBforeignlanguage{en}{Machine
  Learning}}, vol.~53, no.~1, pp. 71--109, Oct. 2003. [Online]. Available:
  \url{https://doi.org/10.1023/A:1025619426553}
\BIBentrySTDinterwordspacing

\bibitem{sewell2008ensemble}
O.~Sagi and L.~Rokach, ``\BIBforeignlanguage{en}{Ensemble learning: {A}
  survey},'' \emph{\BIBforeignlanguage{en}{WIREs Data Mining and Knowledge
  Discovery}}, vol.~8, no.~4, p. e1249, 2018.

\bibitem{ferrari2012using}
A.~Ferrari and S.~Gnesi, ``Using collective intelligence to detect pragmatic
  ambiguities.''\hskip 1em plus 0.5em minus 0.4em\relax IEEE, Sep. 2012, pp.
  191--200.

\bibitem{yang2011analysing}
H.~Yang, A.~De~Roeck, V.~Gervasi, A.~Willis, and B.~Nuseibeh, ``Analysing
  anaphoric ambiguity in natural language requirements,'' \emph{Requirements
  Engineering}, vol.~16, no.~3, pp. 163--189, Sep. 2011.

\bibitem{khan2016survey}
W.~Khan, A.~Daud, J.~A. Nasir, and T.~Amjad, ``A survey on the state-of-the-art
  machine learning models in the context of {NLP},'' \emph{Kuwait Journal of
  Science}, vol.~43, no.~4, Nov. 2016.

\bibitem{vogelsang_automatic_2021}
A.~Vogelsang, D.~Mendez, and M.~Unterkalmsteiner, ``Automatic detection of
  causality in requirement artifacts: the cira approach,'' in
  \emph{Requirements {Engineering}: {Foundation} for {Software} {Quality}: 27th
  {International} {Working} {Conference}, {REFSQ} 2021}, vol. 12685.\hskip 1em
  plus 0.5em minus 0.4em\relax Springer Nature, 2021.

\bibitem{asghar2016automatic}
N.~Asghar, ``Automatic extraction of causal relations from natural language
  texts: A comprehensive survey.''\hskip 1em plus 0.5em minus 0.4em\relax arXiv
  preprint arXiv:1605.07895, 2016.

\bibitem{devlin2018bert}
\BIBentryALTinterwordspacing
J.~Devlin, M.-W. Chang, K.~Lee, and K.~Toutanova, ``{BERT}: {Pre}-training of
  {Deep} {Bidirectional} {Transformers} for {Language} {Understanding},''
  \emph{arXiv:1810.04805 [cs]}, May 2019, arXiv: 1810.04805. [Online].
  Available: \url{http://arxiv.org/abs/1810.04805}
\BIBentrySTDinterwordspacing

\bibitem{joulin2016fasttext}
\BIBentryALTinterwordspacing
A.~Joulin, E.~Grave, P.~Bojanowski, M.~Douze, H.~Jégou, and T.~Mikolov,
  ``{FastText}.zip: {Compressing} text classification models,''
  \emph{arXiv:1612.03651 [cs]}, Dec. 2016, arXiv: 1612.03651. [Online].
  Available: \url{http://arxiv.org/abs/1612.03651}
\BIBentrySTDinterwordspacing

\bibitem{buyukoz2020analyzing}
B.~Büyüköz, A.~Hürriyetoğlu, and A.~Özgür, ``Analyzing {ELMo} and
  {DistilBERT} on {Socio}-political {News} {Classification},'' in
  \emph{Proceedings of the {Workshop} on {Automated} {Extraction} of
  {Socio}-political {Events} from {News} 2020}.\hskip 1em plus 0.5em minus
  0.4em\relax European Language Resources Association (ELRA), May 2020, pp.
  9--18.

\bibitem{shi2014drift}
Z.~Shi, Y.~Wen, C.~Feng, and H.~Zhao, ``Drift detection for multi-label data
  streams based on label grouping and entropy,'' in \emph{2014 {IEEE}
  {International} {Conference} on {Data} {Mining} {Workshop}}.\hskip 1em plus
  0.5em minus 0.4em\relax IEEE, 2014, pp. 724--731.

\bibitem{pant2019multi}
P.~Pant, A.~S. Sabitha, T.~Choudhury, and P.~Dhingra, ``Multi-label
  classification trending challenges and approaches,'' in \emph{Emerging
  {Trends} in {Expert} {Applications} and {Security}}.\hskip 1em plus 0.5em
  minus 0.4em\relax Springer, 2019, pp. 433--444.

\bibitem{chawla2002smote}
N.~V. Chawla, K.~W. Bowyer, L.~O. Hall, and W.~P. Kegelmeyer, ``{SMOTE}:
  synthetic minority over-sampling technique,'' \emph{Journal of artificial
  intelligence research}, vol.~16, pp. 321--357, 2002.

\bibitem{ferrari2017pure}
A.~Ferrari, G.~O. Spagnolo, and S.~Gnesi, ``Pure: {A} dataset of public
  requirements documents,'' in \emph{2017 {IEEE} 25th {International}
  {Requirements} {Engineering} {Conference} ({RE})}.\hskip 1em plus 0.5em minus
  0.4em\relax IEEE, 2017, pp. 502--505.

\end{thebibliography}

\end{document}